\documentclass[10pt, conference]{IEEEtran}
\ifCLASSINFOpdf
   \usepackage[pdftex]{graphicx}
\else
\fi

\usepackage{mdwmath}
\usepackage{mdwtab}
\usepackage{longtable}
\usepackage{slashbox}
\usepackage{supertabular}
\usepackage{ltablex,booktabs}
\usepackage{blindtext}

\usepackage[hidelinks]{hyperref}
\hypersetup{breaklinks=true}
\begin{document}
%
\title{Service Virtualisation of Internet-of-Things Devices:\\ Techniques and Challenges}

\author{\IEEEauthorblockN{Zeinab Farahmandpour\IEEEauthorrefmark{1},
Steve Versteeg\IEEEauthorrefmark{2},
Jun Han\IEEEauthorrefmark{1}, 
Anand Kameswaran\IEEEauthorrefmark{3}}

\IEEEauthorblockA{\IEEEauthorrefmark{1}Swinburne University of Technology, Melbourne, Australia.
Email: \{zfarahmandpour,jhan\}@swin.edu.au}
\IEEEauthorblockA{\IEEEauthorrefmark{2}CA Technologies, Melbourne, Australia. Email: steve.versteeg@ca.com}
\IEEEauthorblockA{\IEEEauthorrefmark{3}CA Technologies, Plano, TX, USA. Email: anand.kameswaran@ca.com}
}

\maketitle

\begin{abstract}
Service virtualization is an approach that uses virtualized environments to automatically test enterprise services in production-like conditions. Many techniques have been proposed to provide such a realistic environment for enterprise services. The Internet-of-Things (IoT) is an emerging field which connects a diverse set of devices over different transport layers, using a variety of protocols. Provisioning a virtual testbed of IoT devices can accelerate IoT application development by enabling automated testing without requiring a continuous connection to the physical devices. One solution is to expand existing enterprise service virtualization to IoT environments. There are various structural differences between the two environments that should be considered to implement appropriate service virtualization for IoT. This paper examines the structural differences between various IoT protocols and enterprise protocols and identifies key technical challenges that need to be addressed to implement service virtualization in IoT environments. 

\end{abstract}

\begin{IEEEkeywords}
Service Virtualisation; Internet-of-Things; Continuous Delivery; 

\end{IEEEkeywords}

%
\IEEEpeerreviewmaketitle

\section{Introduction}

The Internet-of-Things (IoT) is an emerging field, which connects a diverse set of devices over different transport layers, using a variety of protocols. Gartner predicts that by 2020, IoT elements will be incorporated in more than half of major new business processes and systems \cite{Gartner15}. And yet, there are many challenges to readily deliver IoT systems. As such, there is a pressing need to develop techniques to address these challenges. 

As the IoT continues to emerge, there will be a growing number of software applications communicating with IoT devices. The IoT connected software components and applications can be categorised into tiers (as depicted in Figure 1):

\begin{itemize}[\IEEEsetlabelwidth{Z}]
	\item 	Device gateways (GW): responsible for interfacing directly with an IoT device and providing an API (such as REST) to other applications and services
	\item Monitors and data aggregators which collect data from IoT devices (edge nodes)
	\item Applications and services for managing IoT devices
	\item Analytics engines which data mine aggregated IoT data
	\item End user applications viewable on the web or mobile devices
\end{itemize}

Software developers writing IoT applications face challenges, which can delay the release of their application and affect software quality. In particular, to test their application requires interfacing with IoT devices. This may require the physical devices to be present every time the application is fully tested. Furthermore, IoT protocols are very diverse and fragmented, which makes developing and testing for this widespread set of protocols a challenge.

Continuous Delivery (CD) \cite{humble2010continuous} is the industry best practice for accelerating software delivery and increasing software quality. At its core, this includes automating each step of the development release cycle and bringing production-like conditions to every test phase. Due to the physical nature of IoT devices as well as their diversity, this poses a challenge to automation.

For enterprise software development, service virtualisation \cite{michelsen2012service} has been applied as a means of emulating all the other services on which an application under test depends. The key idea of service virtualisation is to observe and log the network communication between an application under test and each other service that it interacts with in its production environment. These logged network traces can then be used to build an interactive model, called a virtual service, for each dependency service. The virtual service is then deployed in an emulation environment, allowing the application under test to send requests to and receive responses from the virtual service, as if it were communicating with the real service. This facilitates the automated testing of a software application in production-like conditions, as is required for continuous delivery.

\begin{figure}[!t]
\centering
\includegraphics[width=3.2in]{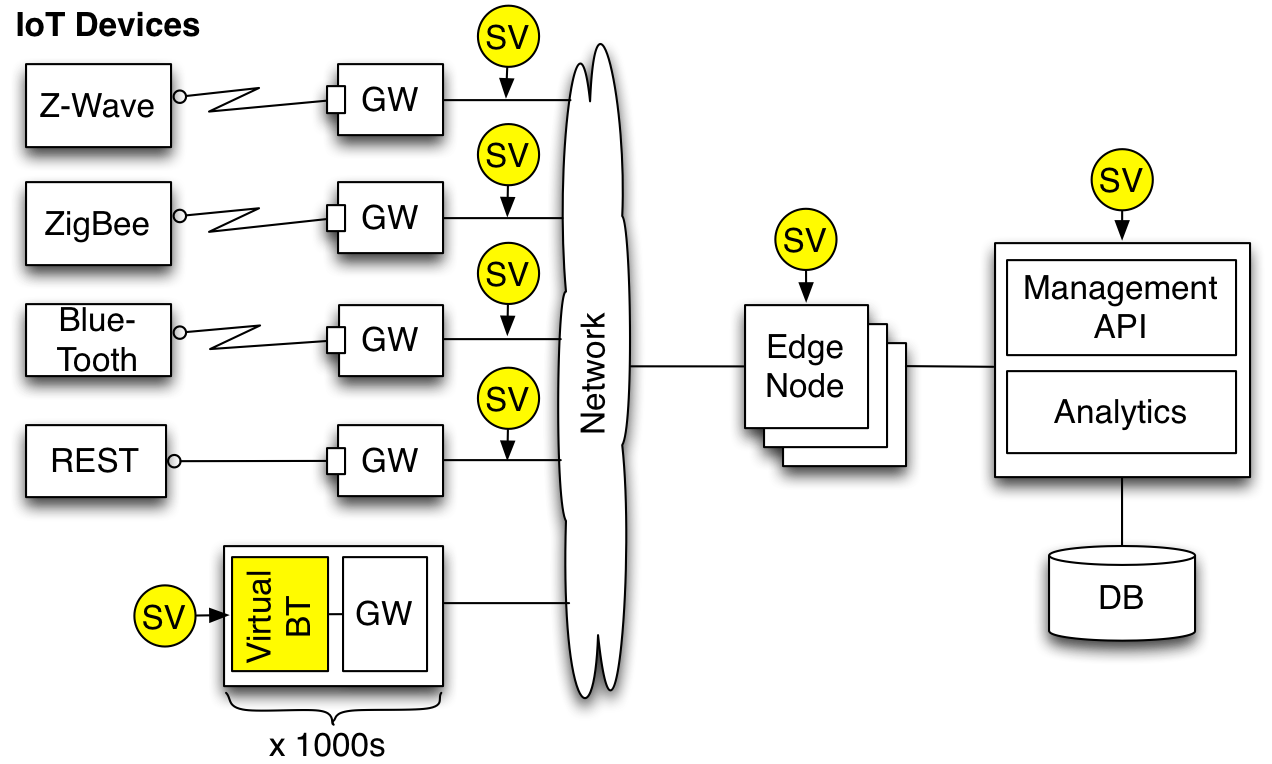}
\caption{Points where service virtualisation (SV) could be applied in IoT}
\label{fig_sim}
\end{figure}

This paper explores whether service virtualisation could be applied to IoT devices to support the realisation of CD for IoT. IoT DevOps problems include heterogeneous hardware, multiple communication layers, lack of industry standards, and skill sets requiring both operations and development. IoT virtualization can remove constraints for IoT solutions development. Provisioning a virtual testbed of IoT devices can accelerate IoT application development by enabling automated testing without requiring a continuous connection to the physical devices. Figure 1 illustrates the points at which IoT applications could be virtualised. In this paper, we survey a sample of IoT protocols to examine their technical differences to enterprise protocols. On this basis, we examine how service virtualisation would need to be adapted to support IoT protocols.

\begin{table*}[ht]
	\caption{IoT Protocol Characteristics}
	\centering

	\resizebox{.9\textwidth}{!}{
	\begin{tabular}[c]{|c||c||c|c|c|c|c|}

 		\hline
		\bf  & \bf & 	\multicolumn{5}{|c|}{\bf   }   \\ 
		\bf  & \bf Enterprise Protocol & 	\multicolumn{5}{|c|}{\bf IoT Protocols  }   \\  \hlx{vh}

		\backslashbox{\bf Taxonomy}{\bf Protocols} & \bf LDAP & \bf MQTT & \bf CoAP & \bf DDS & \bf ZigBee & \bf Z-WAVE \\  \hline	\hlx{vh}
		\hline
		
		\bf UDP/TCP & TCP & TCP & UDP & TCP/UDP &	TCP/UDP & TCP/UDP \\ \hlx{vhv}
	
	\bf Architecture &	Client-server & Pub-sub & Client-server &Pub-sub & Client-server &Client-server \\ 
	\bf  &		&  Client-server\rlap{*}  &  &   Client-server\rlap{*}  &  & \\ \hlx{vhv}
	
	\bf State(ful/less) &	Stateful & Stateful & Stateless	&Stateless&	Configurable	&Configurable \\ \hlx{vhv}

	\bf Communication & Unidirectional\rlap{**} & Unidirectional  &	Unidirectional  & Unidirectional  &	Bidirectional &	Bidirectional \\ 
	
	\bf direction &  & Bidirectional\rlap{*}  &	Bidirectional\rlap{*}  & Bidirectional\rlap{*} &	& \\ \hlx{vhv}

	\bf Header Size	 & Not limited & 2 max 5 bytes & 4 bit fixed header	& 8 bytes&	15 bytes&	Not specified \\ 
	
	\bf  &  &  &  + binary options	& &	& \\ \hlx{vhv}

	\bf coordination &	Asynchronous & Asynchronous&	Asynchronous	&Asynchronous &	Synchronous&	Asynchronous\\ 
	
	\bf  &	Synchronous & &		& Synchronous&	&	\\ \hlx{vhv}

	\bf Network layer &	Application & Session layer&	Session layer	&Session layer	&Sub-application 	& Sub-application \\ 
	
	\bf  &	 & &		&	& (application interface)	&  \\ \hlx{vhv}

	\bf Real-time &	Yes &	No&	No	&Yes&	Yes	&No \\ \hline  
	
	\multicolumn{7}{@{}l}{* IoT protocols have standard and non-standard versions with different structures and properties to meet their environmental needs. Therefore, each}\\

	\multicolumn{7}{@{}l}{ protocol may cover different schemes for each property simultaneously. This highlights the demand for virtualization service environment for IoT.}\\

	\multicolumn{7}{@{}l}{** There is one exception for the LDAP server, which acts as an initiator and can be ignored when the LDAP server sends "Notice of Disconnection" }\\ 
	
	\multicolumn{7}{@{}l}{ to advise the client that the server is going to terminate the LDAP session on its own initiative \cite{sermersheim2006lightweight}.}\\ 
	
	\end{tabular}
	}
\end{table*}

\section{IOT PROTOCOL SURVEY}
To address the many challenges in IoT environments, different standards and communication protocols were introduced. There are a wide range of protocols used by IoT devices. In addition to standardised protocols, there are also many non-standard extensions as well as proprietary protocols. We examine five commonly used IoT protocols which give a spectrum of the potential challenges faced for virtualising an IoT environment. Table 1 summarises some key characteristics of the IoT protocols surveyed. As a comparison point, we also show the attributes of the LDAP protocol \cite{hodges2002lightweight}, as an example enterprise protocol.

\subsection{MQTT}
Message Queue Telemetry Transport (MQTT) \cite{stanford2013mqtt} is a session layer publish-subscribe protocol that is used in applications like the Facebook mobile application. MQTT is an extremely simple and lightweight messaging protocol, designed for constrained devices and low-bandwidth on high-latency or unreliable networks. It is designed to provide embedded connectivity between applications and middleware on one side and networks and communications on the other side. The protocol's architecture consists of three main components: publishers, subscribers, and a broker. Publishers are lightweight sensors that connect to the broker to send their data, then go back to sleep whenever possible. Subscribers are applications that are interested in a certain topic, or a type of sensory data, so that they connect to the broker to be informed whenever new data is received. The broker classifies the sensory data into topics and sends it to interested subscribers.

\subsection{CoAP}
The Constrained Application Protocol (CoAP) \cite{shelby2014constrained} is another session layer protocol that provides a specialized web transfer protocol for use with resource-constrained devices. CoAP is based on the widely successful REST model. Servers make resources available under a URL, and clients access these resources using methods such as GET, PUT, POST, and DELETE. It is built over UDP and has a light-weight mechanism to provide reliability. CoAP contains four messaging modes: confirmable, non-confirmable, piggyback and separate, which support reliable and unreliable transmissions. 

\subsection{DDS}
Data Distribution Service (DDS) \cite{OMGDDS} is a leading data-centric publish-subscribe communication standard. This model builds on the concept of a "global data space" that is accessible to all interested applications. It is a stateless session layer protocol for real-time machine-to-machine communications, that supports both synchronous and asynchronous coordination.

\subsection{ZigBee}
ZigBee \cite{safaric2006zigbee} is a very low-cost, very low-power consumption, two-way, wireless communications standard. Solutions adopting the ZigBee standard are embedded in consumer electronics, building automation, industrial controls, PC peripherals, medical sensor applications, toys, and games. The ZigBee network is comprised of a coordinator, routers and end devices. The coordinator is responsible for initializing, maintaining, and controlling the network. Routers form the network backbone to transfer end devices' packets.

\subsection{Z-WAVE}
The Z-Wave protocol \cite{ZensysZWave} is a low bandwidth half-duplex session layer protocol designed for reliable wireless communication in a low cost control network. The protocol's main purpose is to communicate short control messages in a reliable manner from a control unit to one or more nodes in the network. It follows a master/slave architecture in which the master controls the slaves, sends them commands, and handles and schedules the whole network. It supports an asynchronous architecture communications and is used as a protocol to develop smart products and smart home systems.  

\section{IOT SERVICE VIRTUALISATION CHALLENGES}
Based on the surveyed sample IoT protocols, which are listed in the Table I, we have identified three primary areas where IoT protocols differ from most enterprise protocols, which may pose challenges to implementing service virtualisation for IoT. These include communication challenges, message format challenges and modelling challenges. 

\subsection{COMMUNICATION SYNCRONISATION CHALLENGES}
\subsubsection{Pub/Sub protocols}
IoT protocols such as MQTT and DDS support a Publish/Subscribe architecture. This requires an emulated service to handle situations where a response should be sent in the absence of a triggering request. While service virtualisation has been previously applied to enterprise protocols supporting Publish/Subscribe - it is more difficult to implement than the more widely used client-server protocols, and requires case-by-case implementation. In IoT, Publish/Subscribe architectures are even more prevalent, a generalised approach to emulating Publish/Subscribe therefore requires immediate attention.

\subsubsection{Asynchronous messaging}
As the IoT nodes aim to conserve battery they minimize energy consumption by utilising sleep mode. For example in Z-Wave, at the time when the control node sends a command to a slave node, the slave node may be in sleep mode. Asynchronous communication is therefore adopted to allow messages to be sent at an arbitrary time. For service virtualisation two challenges arise:

\begin{itemize}[\IEEEsetlabelwidth{Z}]
	\item How to correlate requests and responses? 	
	\item How to time when to send responses: this requires keeping track of timestamps.
\end{itemize} 

\subsubsection{Bi-directional communication}
In IoT protocols, the initiation of communication can be unidirectional, bidirectional or a combination of both. For example, in the MQTT protocol, the connection between the publisher and broker as an intervening entity is unidirectional but the connection between the subscriber and broker is bidirectional. Sometimes nodes act as a sender and send their information based on their internal events. For example for a push button, if there is a button press event the node will start sending data without receiving any request. In other situations they respond to "get information" requests from the server, to send their information. The dominant pattern in enterprise protocol service virtualisation, is for service nodes to act as responders, rather than initiators. For the IoT context, the predominant pattern is for bi-directional nodes, capable of acting as both initiators and responders. An IoT service virtualisation solution therefore requires generalised support for bi-directional emulated nodes.

\subsection{MESSAGE FORMAT CHALLENGES}
\subsubsection{Different messaging modes}
There are different modes of messaging in IoT protocols. For example, in the CoAP protocol there are four different messaging modes which can be used based on the requirements, i.e., confirmable, non-confirmable, piggyback and separate. For each mode the structure of response packets is different.

\subsubsection{Chained commands}
A Z-Wave message can contain multiple commands in one message. This is in contrast to most enterprise protocols which have one operation per request. For service virtualisation, this increases the complexity of  message format identification, as chained commands would first need to be separated before they can be processed.

\subsubsection{Fields with less than one byte long}
Since IoT deals with resource constrained devices, protocols try to use shorter packets for their communication. It is more common to have bit fields in IoT compared to in enterprise protocols. This increases the challenge of format identification since fields are not limited by byte boundaries.

\subsection{MODELLING CHALLENGES}

\subsubsection{Encapsulated sensory data}
Sensory data is encapsulated in multiple protocol layers. For example, ZigBee acts as a transport protocol. Application protocols, layered above it, contain the actual sensory information. The sensory data constitutes the key payload information which would need to be captured by any useful virtual service model. However, extracting the sensory data fields from the multiple protocol layers poses a challenge.

\subsubsection{Correlation of data models}
An IoT service virtualization approach needs to provide an accurate simulation of sensory values. It is important to develop and test an appropriate control system to deal with real devices. Therefore service virtualization needs to derive "good" emulation of data coming from sensors in an IoT environment, i.e., the generated data from emulated nodes or sensors should be realistic to help testing the controller. A key question is how close the emulated data should be to a real data stream. Not only do the sensory values need to be accurate, but also sensory values need to be responsive to commands of a controller. For example, for an air-conditioner use case, if we want to generate a model of the temperature sensor, we cannot consider it as an isolated node, because its value is dependent on the commands that the controller sends to the air conditioner. If the temperature that is captured by the sensor is above a threshold, the controller sends a command to the air-conditioner to increase the power of the air conditioner. In response, the temperature is expected to decrease. Therefore for the purpose of service virtualisation, correlation between different elements of the network should be considered in extracting and generating a data model. While this issue also exists in enterprise systems, it is even more paramount in IoT.

\section{IOT OPAQUE SERVICE VIRTUALISATION}
From our analysis it is clear that there are differences between IoT protocols and enterprise protocols which require adaptations to service virtualisation for it to be applied successfully to IoT devices. As identified the key challenges are the large diversity of protocols, communication challenges, message format challenges and data modelling.

Most service virtualisation approaches decode incoming requests into tokens in order to extract fields and values. A set of defined rules based on these fields and values will then be applied to construct a response to send back to the system under test. A limitation of this approach is that it requires a decoder and protocol handler for every protocol. Due to the large diversity and heterogeneity of IoT protocols it is unrealistic to develop and support a protocol handler for every IoT protocol and their variations. This leads us to exploit methods, which use nor or less prior knowledge and try to extract the model of the services automatically.

A recently proposed approach is opaque service virtualization \cite{versteeg2016opaque}. Opaque service virtualisation utilises sequence alignment and data mining methods to analyse samples of recorded messages. Rules for constructing responses are automatically derived. Opaque service virtualisation can be applied to a wide variety of protocols without requiring an individual protocol handler for each protocol. An adaptation of opaque service virtualisation therefore seems well suited to handling the heterogeneity challenge of IoT protocols.

Extensions to opaque service virtualisation are required to handle the communication challenges and the message format challenges. The key consideration is the data modelling challenge. For many enterprise use cases, the data is defined by a schema and is discrete. Protocol operations allow records to be created, read, updated or deleted (CRUD). From a service modelling point of view this is relatively straight forward. Either the record is there or it is not, many of the precise values of the record do not matter for the purpose of the testing scenario. In contrast, for IoT scenarios the continuous nature of the data is integral to the testing scenario (such as a for a controller). For example, as discussed in section C, a temperature sensor has a continuous range of values which is a function of the controller settings and the environment. For a realistic IoT virtual service, opaque service virtualisation needs to be extended to include an explicit data modelling step. Data mining methods could be employed to automatically derive correlations between controller settings and sensory fields. Figure 2 illustrates the conceptual virtual service models with data models included.

\begin{figure}[!t]
	\centering
	\includegraphics[width=3.2in]{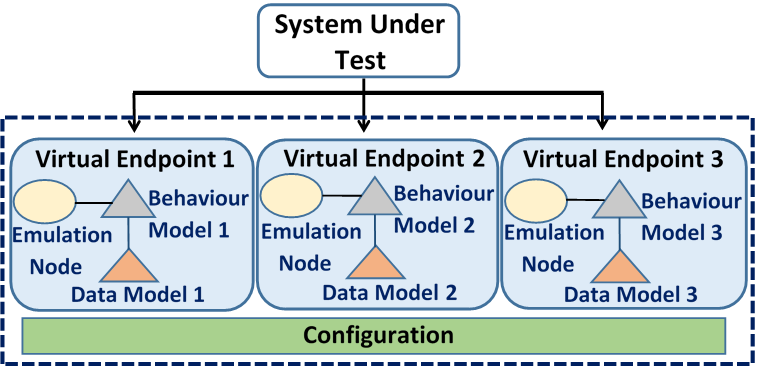}
	\caption{Virtual Testing Environment (VTE)}
	\label{fig_sim}
\end{figure}

\section{CONCLUSION AND FUTURE WORK}
In this paper, we have compared different IoT protocols and enterprise protocols with the focus on expanding service virtualization to the IoT environment. Some key challenges identified for virtualising IoT environments include: heterogeneity, communication synchronisation, formatting complexity and data of continuous nature. To address heterogeneity we plan to implement an extension of Opaque Service Virtualisation. In particular, an explicit data modelling phase needs to be included in the approach. This will allow the automatic virtualisation of IoT environments without requiring prior knowledge of the IoT protocols. We believe this will greatly support IoT developers in enabling them to continuously test their IoT applications in an automated fashion without requiring access to the physical devices.



%





\bibliography{Bibliography}{}
\bibliographystyle{IEEEtran}

\end{document}